\documentclass[aps,pra,twocolumn,10pt,letterpaper,superscriptaddress]{revtex4-1}
\usepackage{amsmath, amssymb}
\usepackage{xcolor, graphicx, ulem}
\usepackage[colorlinks=true,urlcolor=blue,citecolor=blue,linkcolor=blue]{hyperref}

\begin{document}

\title{Quantum coherences of indistinguishable particles}

\author{Jan Sperling}\email{jan.sperling@physics.ox.ac.uk}
\affiliation{Clarendon Laboratory, University of Oxford, Parks Road, Oxford OX1 3PU, United Kingdom}

\author{Armando Perez-Leija}
\affiliation{Max-Born-Institut, Max-Born-Stra\ss{}e 2A, 12489 Berlin, Germany}
\affiliation{Humboldt-Universit\"at zu Berlin, Institut f\"ur Physik, AG Theoretische Optik \& Photonik, 12489 Berlin, Germany}

\author{Kurt Busch}
\affiliation{Max-Born-Institut, Max-Born-Stra\ss{}e 2A, 12489 Berlin, Germany}
\affiliation{Humboldt-Universit\"at zu Berlin, Institut f\"ur Physik, AG Theoretische Optik \& Photonik, 12489 Berlin, Germany}

\author{Ian A. Walmsley}
\affiliation{Clarendon Laboratory, University of Oxford, Parks Road, Oxford OX1 3PU, United Kingdom}

\date{\today}

\begin{abstract}
	We study different notions of quantum correlations in multipartite systems of distinguishable and indistinguishable particles.
	Based on the definition of quantum coherence for a single particle, we consider two possible extensions of this concept to the many-particle scenario and determine the influence of the exchange symmetry.
	Moreover, we characterize the relation of multiparticle coherence to the entanglement of the compound quantum system.
	To support our general treatment with examples, we consider the quantum correlations of a collection of qudits.
	The impact of local and global quantum superpositions on the different forms of quantum correlations is discussed.
	For differently correlated states in the bipartite and multipartite scenarios, we provide a comprehensive characterization of the various forms and origins of quantum correlations.
\end{abstract}

\maketitle

\section{Introduction}

	Since its discovery, the quantum superposition principle has been used to explain a number of observations which are incompatible with classical physics \cite{S35}.
	Such superpositions lead to quantum interferences which, for example, confirm the wave nature of electrons \cite{SMP73}.
	Even the phenomenon of entanglement can be assigned to nonlocal superpositions of quantum states \cite{EPR35}.
	In addition, quantum treatment of multiple indistinguishable particles yields wave functions which have to satisfy an exchange symmetry, also leading to superpositions.
	That is, the joint state of bosons or fermions is a symmetric or antisymmetric linear combination, respectively, of all permutations of the single-particle states \cite{MG64,S94}.
	Aside from their fundamental character, quantum interferences also result in resources which can be exploited for quantum technologies \cite{DM03,SAP16}.
	For instance, the basic carrier of a bit of quantum information is the qubit, which can be an arbitrary superposition state beyond classical truth values \cite{NC00}.
	Hence, the characterization of quantum interferences is vital to get a deeper insight into quantum phenomena and their practical application.

	To compare classical and quantum characteristics of a system, the concept of a classical state has to be specified to classify the resources for quantum protocols \cite{SAP16,NC00}.
	One approach to achieve this is the introduction of the notion of quantum coherence \cite{BCP14,LM14}.
	The main idea is that one identifies reference states which have properties that are compatible with a classical model.
	Whenever the state is not in the family of classical states, it is incompatible with the considered classical model and, thus, certifies some quantum characteristics.
	For example, coherent states, as introduced by Schr\"odinger, were shown to exhibit a resemblance to a classical harmonic oscillator \cite{S26}.
	This had a remarkable impact on the development of quantum optics \cite{S63,G63,MW65} and has been further generalized to nonlinear scenarios \cite{P72,MV96,MMSZ97}.
	The classification of the quantum coherence of a system has recently gained a lot of attention since it describes a rather general approach and can be applied to many physical scenarios.
	To quantify the quantumness of a system, different quantum coherence measures have been introduced (see, for example, Refs. \cite{BCP14,LM14,SV15,WY16}).
	It is worth pointing out that for an unambiguous quantification, the importance of superpositions has been additionally demonstrated \cite{SV15}.

	Different instances of quantum coherences have been compared and analyzed, e.g., in Refs. \cite{MS16,VS17,YDXLS17}.
	For instance, it has been shown that quantum coherence can be used as a resource to generate quantum entanglement \cite{SSDBA15,KSP16}.
	This especially leads to a one-to-one transfer of single-mode nonclassical correlations to entanglement in quantum optics \cite{VS14}.
	Moreover, the joint description of quantum coherence in multipartite systems has been recently studied for the cases in which the subsystems are connected to different classical models \cite{SAWV17}, which was also experimentally applied in Ref. \cite{ASCBZV17}.
	In this context, it was also shown that entanglement is always one form of quantum coherence, independently of the classical model.

	To identify entanglement, various criteria for its detection have been formulated \cite{HHHH09,GT09}.
	Among the most established approaches is the method of entanglement witnesses \cite{HHH96,HHH01}.
	It allows one to certify entanglement by means of Hermitian operators which have specific properties, and it has been applied to different experimental scenarios, e.g., in Refs. \cite{BMNMAM03,RSNK07,FMAVSO12,GPMMLP12}.
	Moreover, techniques have been formulated which allow for the systematic construction and optimization of entanglement witnesses from general observables \cite{SV09,SV13}.
	This technique has been experimentally applied to identify entanglement by means of nonlocal superpositions \cite{GPMMSVP14} or to characterize complex forms of multipartite entanglement \cite{GSVCRTF15,GSVCRTF16}.

	Because of the impact of the exchange symmetry on the formulation of a quantum state, already the definition of entanglement in systems of identical particles is challenging and the subject of an ongoing debate \cite{GMW02,BFT14}.
	Still, it has been shown that entanglement of indistinguishable particles can be transformed into entanglement of distinguishable degrees of freedom \cite{CMMTF07,KCP14}.
	Also, the construction of witnesses has been generalized for studying bosonic and fermionic entanglement \cite{ESBL02,GKM11,IV13,RSV15}.

	Thus, the individual forms of quantum correlations have been frequently studied.
	However, their interplay has not been investigated in a unified framework so far.
	Hence, we aim to address such mutual influences of different forms and origins of quantum coherence.

	In this paper, we perform a joint characterization of quantum coherence, entanglement, and the exchange symmetry of multiparticle quantum systems.
	We compare those notions of quantumness and exploit their relation.
	By constructing a number of comprehensive examples, we discuss the impact of local and global quantum superpositions on the quantumness of the individual subsystems as well as the joint system.
	In particular, the influence of the exchange symmetry of boson and fermions is considered.
	Based on the introduction of a directly accessible witnessing approach for different forms of quantum coherences, we quantify the operational strength of the quantum correlations for a system of qudits.

	The structure of this contribution is as follows.
	In Sec. \ref{sec:Preliminaries}, we give a brief introduction to the methods used in this work, and it is followed by an intermediate classification of the studied forms of quantum coherence.
	In Sec. \ref{sec:MultiparticleQuantumness}, we elaborate two concept of generalizing quantum coherence from the single-particle to the multiparticle scenario, which includes the cases of distinguishable and indistinguishable particles.
	Collections of qudits are characterized and extensively discussed in Sec. \ref{sec:Qudit} to exemplify our general findings.
	We summarize and conclude our work in Sec. \ref{sec:Conclusions}.

\section{Preliminaries}\label{sec:Preliminaries}

	In this section, we recall the concept of incoherent states, which is typically used to define the reference of classical density operators of a single particle.
	We also formulate an approach to probe nonclassical correlations based on observable criteria.
	In the second part of this section, we briefly summarize the treatment of many indistinguishable particles in quantum physics.
	Eventually, we further motivate the main aim of this work.

\subsection{Witnessing quantum coherence}\label{subsec:PrelimSingPart}

	Let us consider a set of pure classical states $\mathcal C$, which is a closed subset of $\mathcal H$.
	The notion ``classical'' can have different meanings when considering different physical scenarios.
	One particular example is specified later.
	For the time being, let us keep this concept as general as possible.
	A convex-hull construction allows us to generalize classical states from pure states, $|c\rangle\in\mathcal C$, to mixed ones \cite{comment2}.
	Including classical statistical correlations, this defines the set of incoherent states $\rho\in\mathcal I$, with
	\begin{align}\label{eq:IncoherentStates}
		\rho=\int_{|c\rangle\in\mathcal C} dP(|c\rangle) |c\rangle\langle c|,
	\end{align}
	where $P$ is a probability distribution; see Ref. \cite{SAP16} for a recent review.
	A state which is not a mixture of pure classical states, $\varrho\notin\mathcal I$, will be denoted as a quantum correlated state throughout this work.
	Such a state $\varrho$ exhibits quantum features which are inconsistent with the concept of classicality under study.

	One way to infer quantum correlations is given in terms of measurable observables $L$.
	For such a bounded, Hermitian operator $L$, one can compute the upper bound of its expectation value for classical states, $g_{\max}=\sup_{\rho\in\mathcal I}\mathrm{tr}[\rho L]$.
	Due to the convexity of $\mathcal I$ and the generation of $\mathcal I$ via the pure states in $\mathcal C$, one can also directly write the bound $g_{\max}$ in the form
	\begin{align}\label{eq:Bounds}
		g_{\max}=\sup_{|c\rangle\in\mathcal C} \langle c|L|c \rangle.
	\end{align}
	Therefore, whenever a state $\varrho$ yields an expectation value $\langle L\rangle=\mathrm{tr}[\varrho L]$ which overcomes this bound,
	\begin{align}\label{eq:Criterion}
		\langle L\rangle>g_{\max},
	\end{align}
	we have certified the quantum correlation of $\varrho$.

	Moreover, it is worth mentioning that for any $\varrho\notin\mathcal I$, there exists an observable $L$ which can identify its quantum correlations based on the above criterion \eqref{eq:Criterion}.
	This is a direct application of the Hahn-Banach separation theorem \cite{Y08}, and it has similar applications.
	For example, such a separation of sets of classical states from quantum correlated ones via Hermitian operators has been discussed in the context of entanglement witnesses \cite{HHH96,HHH01,T05}.
	Here, a witness for the studied forms of quantum correlations can be written as
	\begin{align}
		W=g_{\max}\mathbb I-L,
	\end{align}
	where $\mathbb I$ is the identity.
	The quantum correlation condition \eqref{eq:Criterion} implies that for all classical states $\langle W\rangle \geq 0$ holds.
	Consequently, $\langle W\rangle<0$ uncovers quantum coherence.
	For instance, if the classical states are considered to be coherent states of an optical mode, one obtains witnesses to probe nonclassical light \cite{MSVH14}.

	As already mentioned, a similar method was used to construct entanglement criteria \cite{SV09,SV13,RSV15}.
	In this scenario, the set $\mathcal I$ of classical states includes all separable states \cite{W89}, $\mathcal I=\mathcal I^\mathrm{(sep)}$.
	The underlying optimization problem that has to be solved to obtain the corresponding bound \eqref{eq:Bounds} for separable states, $g_{\max}=g_{\max}^\mathrm{(sep)}$, was derived in Refs.  \cite{SV09,SV13,RSV15} and is also briefly presented in the Appendix.
	This method is used in this work for identifying entanglement.

	For the verification of quantum coherence of a pure state $|\psi\rangle$ ($\langle \psi|\psi\rangle=1$), one can alternatively compute the fidelity with classical states,
	\begin{align}
		F(|\psi\rangle)=\sup_{|c\rangle\in\mathcal C}|\langle c|\psi\rangle|.
	\end{align}
	The fidelity is one, $F(|\psi\rangle)=1$, if and only if $|\psi\rangle\langle\psi|\in\mathcal I$ (equivalently, $|\psi\rangle\in\mathcal C$).
	If the fidelity is smaller than one, $F(|\psi\rangle)<1$, we have $|\psi\rangle\langle\psi|\notin\mathcal I$.
	In terms of the condition \eqref{eq:Criterion}, we can therefore define the test operator
	\begin{align}
		L=|\psi\rangle\langle \psi|.
	\end{align}
	In this case, we get the expectation value $\langle \psi|L|\psi\rangle=1$, and the bound $g_{\max}$ is given by
	\begin{align}
		g_{\max}=F(|\psi\rangle)^2.
	\end{align}
	Thus, we have $\langle L\rangle=1>g_{\max}=F(|\psi\rangle)^2$ if and only if the state $|\psi\rangle$ is quantum correlated, $|\psi\rangle\notin\mathcal C$.
	Also note that such a rank-one operator $L$  can be applied to arbitrary states $\varrho$ as well, which yields the sufficient quantum correlation condition $\langle \psi|\varrho|\psi\rangle>F(|\psi\rangle)^2$.

\subsection{Multiparticle description}

	In order to describe two- and multiparticle systems, we briefly recapitulate the standard notation of multipartite systems \cite{S94}.
	For multiple distinguishable particles, one applies the tensor product $\otimes$ to describe the joint system.
	For example, $|a\rangle\otimes|b\rangle\in\mathcal H^{\otimes 2}$ describes a two-particle product state without any exchange symmetry.

	In the case of two bosons, one uses symmetric tensor-product states.
	Such states are defined through
	\begin{align}\label{eq:TensorBoson}
		|a\rangle\vee|b\rangle\cong|a\rangle\otimes|b\rangle+|b\rangle\otimes|a\rangle\in\mathcal H^{\vee 2}\subset\mathcal H^{\otimes 2}.
	\end{align}
	We use the symbol ``$\cong$'' to indicate that a proper normalization has to be performed; that is, the left- and right-hand sides are identical up to a normalization constant.
	Note that we can also have $|a\rangle\vee|a\rangle\cong|a\rangle\otimes|a\rangle$.
	This type of symmetric tensor product $\vee$ is used to describe indistinguishable particles, which have a symmetric (i.e., positive) exchange symmetry.

	For fermions, the skew-symmetric tensor product $\wedge$ is applied.
	For instance, a bipartite state can be written as
	\begin{align}\label{eq:TensorFermion}
		|a\rangle\wedge|b\rangle\cong|a\rangle\otimes|b\rangle-|b\rangle\otimes|a\rangle\in\mathcal H^{\wedge 2}\subset\mathcal H^{\otimes 2}.
	\end{align}
	Here, we have $|a\rangle\wedge|a\rangle=0$ (the null vector).
	Identifying $|a\rangle$ and $|b\rangle$ with wave functions, the skew-symmetric tensor product \eqref{eq:TensorFermion} relates to the well-known Slater determinant, describing the antisymmetric (i.e., negative) exchange symmetry of fermions.

	For the general $N$-mode case, it is convenient to introduce an (anti)symmetrization projection operator,
	\begin{align}\label{eq:PermutationOperator}
		\Pi_{\pm}=\frac{1}{N!}\sum_{\sigma\in\mathcal S_N}(\pm1)^{|\sigma|} P_{\sigma},
	\end{align}
	where $\mathcal S_N$ is the permutation group and $(-1)^{|\sigma|}$ is the parity of the permutation $\sigma$.
	Each permutation operator $P_\sigma$ exchanges the subsystems according to $\sigma\in\mathcal S_N$, which means $P_\sigma[|x_1\rangle\otimes\cdots\otimes|x_N\rangle]=|x_{\sigma[1]}\rangle\otimes\cdots\otimes|x_{\sigma[N]}\rangle$.
	Hence, for the two-particle example, the product states in Eqs. \eqref{eq:TensorBoson} and \eqref{eq:TensorFermion} can be alternatively written as
	\begin{align}
	\begin{aligned}
		|a\rangle\vee|b\rangle\cong&\Pi_{+}[|a\rangle\otimes|b\rangle]
		\\\text{ and }
		|a\rangle\wedge|b\rangle\cong&\Pi_{-}[|a\rangle\otimes|b\rangle],
	\end{aligned}
	\end{align}
	respectively.

\subsection{Origins of quantum correlations}

\begin{figure}[b]
	\includegraphics[width=5.5cm]{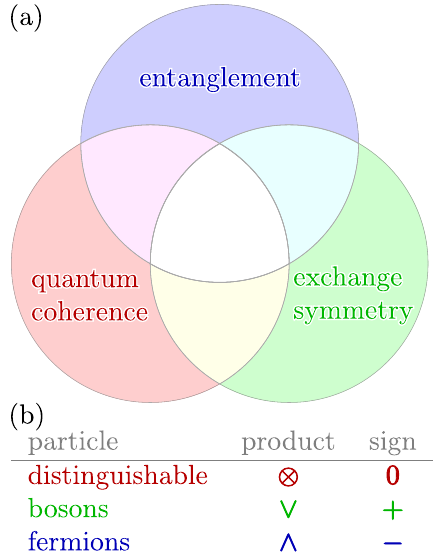}
	\caption{(Color online)
		The interplay of different origins of multiparticle quantum correlations is schematically outlined in (a).
		The considered types of particles together with the underlying tensor products and the signs used for their identification are listed in (b).
	}\label{fig:Concept}
\end{figure}

	In this work, we characterize quantum superposition effects, i.e., quantum coherences, which can have different origins (see Fig. \ref{fig:Concept}).
	In a single-mode system, we can have quantum superpositions of pure classical states, such as
	\begin{align}
		|\psi\rangle\cong|c_1\rangle+|c_2\rangle,
	\end{align}
	for linearly independent $|c_1\rangle,|c_2\rangle\in\mathcal C$.
	Besides local superposition, we might also have global superpositions in multipartite systems of distinguishable particles, which resemble the phenomenon of entanglement \cite{SAWV17}.
	Also, the exchange symmetry of indistinguishable particles can be another reason for the occurrence of superpositions [see Eqs. \eqref{eq:TensorBoson} and \eqref{eq:TensorFermion}].

	Hence, our objective is to identify the possible source of quantum superpositions, which are in our case the aforementioned local superpositions, entanglement, and the exchange symmetry.
	This is achieved by detecting quantum interferences based on the previously formulated test criteria.
	To discern between the cases of distinguishable particles, bosons, and fermions, we use the superscripts $(0)$, $(+)$, and $(-)$, respectively, to indicate which notion is studied (see Fig. \ref{fig:Concept}).

\section{Quantum coherence in multiparticle systems}\label{sec:MultiparticleQuantumness}

	Now, let us study how to generalize the concept of quantum correlations from a single particle to multiple particles.
	In the first step, distinguishable particles are considered before studying bosons and fermions.
	Further on, the relation to entanglement is considered.
	This eventually leads to a discussion of the known problem of how to define entanglement for indistinguishable particles and its generalization to quantum coherence.
	Finally, the quantum correlations are specified for uncovering the different types of quantum effects.

\subsection{Distinguishable particles}

	Multimode incoherent states are elements of the closure (with respect to the trace norm) of the convex span of product incoherent states,
	\begin{align}
		\mathcal I^{(0)}=\overline{\mathrm{conv}\{\rho_1\otimes\cdots\otimes\rho_N:\rho_j\in\mathcal I_j\text{ for } j=1,\ldots,N\}},
	\end{align}
	for given notions of single-particle incoherent states $\rho_j\in\mathcal I_j$ for the subsystems $j=1,\ldots,N$.
	Equivalently, one could start with the pure states and define \cite{BCA15}
	\begin{align}\label{eq:TensorClassical}
		\mathcal C^{(0)}=\{|c_1\rangle\otimes\cdots\otimes|c_N\rangle: |c_j\rangle\in\mathcal C_j\text{ for }j=1,\ldots,N\},
	\end{align}
	which leads to the same set $\mathcal I^{(0)}$ of incoherent and possibly mixed states.
	Let us stress that the superscript (0) indicates that distinguishable particles are currently considered.
	For the following generalization of this concept to indistinguishable particles, we assume that the individual concepts of classical states are identical for all subsystems,
	\begin{align}
		\mathcal C_1=\cdots=\mathcal C_N=\mathcal C.
	\end{align}
	Again, any state $\varrho$ which is not a statistical mixture of classical states, $\varrho\notin\mathcal I^{(0)}$, refers to a (multiparticle) quantum correlated state.

	As we will demonstrate, quantum coherence can stem from local quantum correlations, which have their origin in single-particle nonclassicality, as well as from global quantum correlations, requiring a multimode description.
	In particular, entangled states are examples of quantum correlated states which include global quantum correlations \cite{SAWV17}.
	To show this, we can make the simple observation that the definition \eqref{eq:TensorClassical} implies that $\mathcal C^{(0)}$ solely includes $N$-particle tensor-product states.
	If $N=2$ and $|a\rangle,|b\rangle\in\mathcal C$ are linearly independent classical states, we can therefore conclude that
	\begin{align}\label{eq:EntTensorGeneral}
		|\phi_{\varphi}\rangle\cong|a\rangle\otimes|b\rangle+e^{i\varphi}|b\rangle\otimes|a\rangle
	\end{align}
	is two particle quantum correlated and that this nonclassicality is due to entanglement for all phases, $0\leq\varphi<2\pi$.

\subsection{Indistinguishable particles}

	As mentioned earlier, the (anti)symmetrization, employed to describe (fermions) bosons, can be obtained by applying the permutation operator $\Pi_{\pm}$ in Eq. \eqref{eq:PermutationOperator}.
	Therefore, the symmetric and skew-symmetric generalizations of multimode classical states can be given as
	\begin{align}\label{eq:AlgebraClassical}
		\mathcal C^{(\pm)}\cong\Pi_{\pm}\mathcal C^{(0)}.
	\end{align}
	The mixed, incoherent states in systems of indistinguishable particles are consequently labeled as $\mathcal I^{(\pm)}$.
	At first sight, this definition seems to be the natural extension of the concept of quantum coherence from distinguishable particles to indistinguishable ones.
	However, let us critically reconsider this, which was analogously done for the different notions of entanglement of bosons and fermions \cite{BFT14}.

	Applying the definition \eqref{eq:AlgebraClassical}, the states $|a\rangle\vee|b\rangle$ and $|a\rangle\wedge|b\rangle$ turn out to be classical for any linearly independent pair $|a\rangle,|b\rangle\in\mathcal C$.
	Yet, for the case of distinguishable particles and in connection with the state in Eq. \eqref{eq:EntTensorGeneral} for $e^{i\varphi}\in\{+1,-1\}$, we also have discussed that these states are quantum correlated.
	Thus, we can say that the question of existing multiparticle quantum correlations for a given quantum state can have different answers, depending on how the joint system is composed.
	On the one hand, this means that depending on whether or not the exchange symmetry is included in the definition of the multiparticle classical reference, we find that
	\begin{enumerate}
		\item[(i)] the states $|a\rangle\vee|b\rangle$ and $|a\rangle\wedge|b\rangle$ are classical with respect to $\mathcal C^{(\pm)}$,
		\item[(ii)] or $|a\rangle\otimes|b\rangle\pm|b\rangle\otimes|a\rangle$ are quantum correlated with respect to $\mathcal C^{(0)}$,
	\end{enumerate}
	even though they describe the same state.
	On the other hand, this difference can be a helpful tool to identify superpositions stemming from the exchange symmetry.
	For instance, if we have a state $\varrho$ for which $\varrho\notin\mathcal I^{(0)}$ and $\varrho\in\mathcal I^{(\pm)}$ hold, we can conclude that interferences are a consequence of indistinguishability.

\subsection{Entanglement}\label{subsec:QuestionEntanglement}

	As a first, trivial example, we consider the case that all single-particle states are classical, $\mathcal C\cong\mathcal H$.
	Then, definition \eqref{eq:TensorClassical} implies that the incoherent states are exclusively the separable ones and entangled states are quantum correlated \cite{W89}.
	Even in this simplified case, which cannot exhibit local (i.e., single-particle) quantum correlations, we can have global quantum correlations (i.e., entanglement) \cite{SAWV17,SLR17}.

	Again, the state \eqref{eq:EntTensorGeneral} is entangled with respect to the tensor product, $|\phi_{\varphi}\rangle\notin\mathcal C^{(0)}$ for $e^{i\varphi}=\pm1$, using the standard definition of entanglement \cite{W89}.
	Using $\vee$ and $\wedge$ as a basis to define entanglement \cite{OK13,GKM11}, these state are separable, $|\phi_{0}\rangle\in\mathcal C^{(+)}$ and $|\phi_{\pi}\rangle\in\mathcal C^{(-)}$.
	In the latter interpretation, the exchange symmetry of bosons and fermions cannot lead to entanglement.

	Hence, depending on option (i) or (ii), we do not or do have quantum correlations, even for our trivial choice that all single-particle states are classical, $\mathcal C\cong\mathcal H$.
	Both concepts, the state \eqref{eq:EntTensorGeneral} for $\varphi=0,\pi$ is either inseparable (using $\otimes$) or separable (using $\vee$ or $\wedge$), have been discussed, used, and compared in the literature \cite{BFT14,AMN07}.
	The question of which of those interpretations is correct is still open, and it might not even have a definitive answer.
	Here, we compare both possibilities and draw the corresponding conclusions for each of the notions independently, without any personal preference.

	Besides options (i) and (ii), we can ask ourselves if the exchange symmetry can be used to generate quantum correlations.
	Or can superpositions stemming from the exchange symmetry be activated?
	In Refs. \cite{CMMTF07,KCP14}, it was already shown that superpositions originating from the (anti)symmetrization can be converted into entanglement of distinguishable degrees of freedom.
	Since entanglement is always one form of quantum correlations for all sets $\mathcal C$ of single-mode classical states and distinguishable particles, one can conclude that superpositions resulting from the exchange symmetry can be (at least) converted into a physically relevant form of quantum coherences.
	Furthermore, it is noteworthy that any local form of quantum coherence can be transformed into entanglement \cite{KSP16}.

\subsection{Unified witnessing of different origins of quantum correlations}

	In order to probe which quantum effect might be present in a given state, we now formulate unified criteria.
	We can apply the general condition \eqref{eq:Criterion}, including the bound \eqref{eq:Bounds}, for the different origins of quantum correlations.
	Such criteria for a given observable $L$ read
	\begin{align}\label{eq:UnifiedCrierion}
	\begin{aligned}
		\langle L\rangle&>g^{(0)}_{\max},
		\\
		\langle L\rangle&>g^{(\pm)}_{\max},
		\\
		\langle L\rangle&>g^{(\mathrm{sep},0)}_{\max},
		\\
		\langle L\rangle&>g^{(\mathrm{sep},\pm)}_{\max}.
	\end{aligned}
	\end{align}
	Here, the bounds determine the quantum correlation under consideration.
	The criterion which includes the bound $g^{(0)}_{\max}$ probes general quantum coherence regardless of the exchange symmetry.
	The criterion based on the bound $g^{(\pm)}_{\max}$ verifies quantum coherence and excludes effects originating from the exchange symmetry.
	When we want to exclude local coherence, we can detect global quantum coherences with the criterion using $g^{(\mathrm{sep},0)}_{\max}$; it verifies inseparability for distinguishable degrees of freedom.
	Finally, entanglement which is not a result of the symmetrization postulate for bosons or fermions is certified with the condition applying the bound $g^{(\mathrm{sep},\pm)}_{\max}$.

	Hence, the quantum correlation tests \eqref{eq:UnifiedCrierion} allow for the unified determination of quantum effects.
	Moreover, they also enable us to infer the origin of quantum correlations with a single observable.
	For instance, if quantum correlations are verified by the first condition, $\langle L\rangle>g^{(0)}_{\max}$, and no entanglement is detected, $\langle L\rangle \leq g^{(\mathrm{sep},0)}_{\max}$, we can conclude that the verified forms of quantum correlations in terms of the operator $L$ describe local effects and not global ones.
	Another example could be the successful identification of quantum correlations in the sense of distinguishable particles, $\langle L\rangle >g^{(0)}_{\max}$, but for indistinguishable particles, the test fails, $\langle L\rangle\leq g^{(\pm)}_{\max}$.
	This means that the detected quantum coherence originates from the exchange symmetry.

	Let us emphasize that $\langle L\rangle\leq g_{\max}$ does not exclude quantum correlations.
	Another choice of operator, say $L'$, might lead to $\langle L'\rangle> g'_{\max}$.
	Hence, we consistently speak about the detectable quantum correlations in terms of a given measurement $L$.
	Still, we already pointed out earlier that quantum correlations of a pure state $|\psi\rangle$ are present if and only if they can be verified in terms of $L=|\psi\rangle\langle\psi|$.

\section{Application: Qudits}\label{sec:Qudit}

	In the previous section, we compared the notion of quantum coherence in multipartite systems of distinguishable particles as well as bosons and fermions in general.
	In this section, we apply the previously formulated concepts to a specific example.
	States are constructed such that the different interpretations of quantum correlations in multiparticle systems lead to different conclusions.
	Such considerations also clarify some rather abstract points mentioned previously.

	The scenario we are going to study is described via a $d$-dimensional, single-particle Hilbert space $\mathcal H$.
	In a classical $d$-level system, a pure classical particle can take one of the integer values $k=0,\ldots,d-1$.
	Thus, the following orthonormal basis of $\mathcal H$ defines the set of pure classical states:
	\begin{align}\label{eq:SimpleClassical}
		\mathcal C=\{|k\rangle: k=0,\ldots,d-1\}.
	\end{align}
	Statistical fluctuations allow for describing the incoherent mixed states in $\mathcal I$ of this qudit system via Eq. \eqref{eq:IncoherentStates},
	\begin{align}
		\rho\in\mathcal I
		\Leftrightarrow
		\rho=\sum_{k=0}^{d-1} p_k|k\rangle\langle k|,
	\end{align}
	for $p_k\geq 0$ and $\sum_k p_k=1$.

	For the following investigations, it is worth pointing out that $\dim\mathcal H^{\otimes N}=d^N$, $\dim\mathcal H^{\vee N}=\binom{N+d-1}{N}$, and $\dim\mathcal H^{\wedge N}=\binom{d}{N}$.
	Especially for the latter case, we get zero dimensionality for $d<N$.
	Thus, we assume for convenience that $d$ is sufficiently large.
	In principle, $\mathcal H$ could also be an infinite-dimensional system, i.e., $d=\infty$.

\subsection{Incoherent multi-qudit states}

	In order to formulate the incoherent density operators for multiple qudits, we need to formulate the pure, classical, and normalized states by applying the different definitions for distinguishable particles [Eq. \eqref{eq:TensorClassical}] and indistinguishable ones [Eq. \eqref{eq:AlgebraClassical}].
	For the former case, we directly get the elements of the set $\mathcal C^{(0)}$ as
	\begin{align}
		|k_1\rangle\otimes\cdots\otimes|k_N\rangle,
		\\\nonumber
		\text{ for }
		0\leq k_j\leq d-1
		\text{ and }
		j=1,\ldots,N.
	\end{align}
	For antisymmetric fermions, we find that all normalized states in $\mathcal C^{(-)}$ take the form
	\begin{align}
		\sqrt{N!}\Pi_{-}\left[ |k_1\rangle\otimes\cdots\otimes|k_N\rangle \right]
		=\frac{|k_1\rangle\wedge\cdots\wedge|k_N\rangle}{\sqrt{N!}},
		\\\nonumber
		\text{for }0\leq k_1<\ldots<k_N\leq d-1.
	\end{align}
	An example is the bipartite ($N=2$), skew-symmetric product state $|0\rangle\wedge|1\rangle/\sqrt{2!}=(|0\rangle\otimes|1\rangle-|1\rangle\otimes|0\rangle)/\sqrt{2}$.
	Analogously, for the symmetric case of bosons, the set $\mathcal C^{(+)}$ is completely determined by the normalized states
	\begin{align}
		&\sqrt{\frac{N!}{N_0!\cdots N_{d-1}!}}\Pi_{+}\left[ |k_1\rangle\otimes\cdots\otimes|k_N\rangle \right]
		\\\nonumber
		=&\frac{|k_1\rangle\vee\cdots\vee|k_N\rangle}{\sqrt{N!N_0!\cdots N_{d-1}!}},
		\text{ for }0\leq k_1\leq\ldots\leq k_N\leq d-1,
	\end{align}
	where we additionally define $N_k$, which counts how often a given value $k\in\{0,\ldots, d-1\}$ occurs in the tuple $(k_1,\ldots,k_N)$,
	\begin{align}
		N_k=|\{k_j:k_j=k\text{ for }j=1,\ldots,N\}|.
	\end{align}
	A tripartite ($N=3$) example is $|0\rangle\vee|0\rangle\vee|1\rangle/\sqrt{3!2!1!}=(|0\rangle\otimes|0\rangle\otimes|1\rangle+|0\rangle\otimes|1\rangle\otimes|0\rangle+|1\rangle\otimes|0\rangle\otimes|0\rangle)/\sqrt{3}$, which is a symmetric tensor product $\vee$ of $N_0=2$ times the state $|0\rangle$ and $N_1=1$ times $|1\rangle$.
	Mixtures of those pure states [see Eq. \eqref{eq:IncoherentStates}] extend the definition of classicality to mixed quantum states in an $N$-particle system of qudits.

\subsection{Detection of multiparticle quantum correlations}\label{subsect:QuditDetection}

	To probe the quantum coherences via the general criterion \eqref{eq:UnifiedCrierion} and applying the technique from Sec. \ref{subsec:PrelimSingPart}, we may expand any operator $L$ as
	\begin{align}
	\begin{aligned}
		L=\sum_{k_1,\ldots,k_N,l_1,\ldots,l_N=0}^{d-1}& L_{(k_1,\ldots,k_N),(l_1,\ldots,l_N)}
		\\
		&\times |k_1\rangle\langle l_1|\otimes\cdots\otimes|k_N\rangle\langle l_N|.
	\end{aligned}
	\end{align}
	Using the above definitions of classical states for the different scenarios of compositions, we get the following bounds for classically correlated states:
	\begin{align}\label{eq:MultiparticleBoundsTensor}
		g_{\max}^{(0)}=&\max_{k_1,\ldots,k_N}\left[ L_{(k_1,\ldots,k_N),(k_1,\ldots,k_N)} \right],
		\\\nonumber
		g_{\max}^{(-)}=&\max_{k_1<\ldots<k_N}\Bigg[
			\frac{1}{N!}
			\sum_{\sigma,\tau\in\mathcal S_N} (-1)^{|\sigma|+|\tau|}
		\\&\times\label{eq:MultiparticleBoundsFermions}
			L_{(k_{\sigma[1]},\ldots,k_{\sigma[N]}),(k_{\tau[1]},\ldots,k_{\tau[N]})}
		\Bigg],
		\\\nonumber
		g_{\max}^{(+)}=&\max_{k_1\leq\ldots\leq k_N}\Bigg[\frac{1}{N!N_0!\cdots N_{d-1}!}
		\\&\times\label{eq:MultiparticleBoundsBosons}
			\sum_{\sigma,\tau\in\mathcal S_N} L_{(k_{\sigma[1]},\ldots,k_{\sigma[N]}),(k_{\tau[1]},\ldots,k_{\tau[N]})}
		\Bigg].
	\end{align}
	Let us stress again that $g_{\max}^{(0)}$ yields quantum correlations in terms of tensor-product states, and $g_{\max}^{(\pm)}$ is the bound for the products which include the corresponding exchange symmetry of bosons and fermions.

\subsection{A single particle}\label{subsec:Qudit1}

	Before considering multiparticle examples, we briefly discuss a single qudit first.
	We directly get the bound of classical states for a given observable $L=\sum_{k,l} L_{k,l}|k\rangle\langle l|$,
	\begin{align}
		g_{\max}=\max_{k=0,\ldots,d-1} L_{k,k}
	\end{align}
	[see also Eq. \eqref{eq:Bounds}].
	For instance, as discussed in Sec. \ref{subsec:PrelimSingPart}, the quantum correlations of a superposition state
	\begin{align}\label{eq:SimpleState}
		|s_{k,l}\rangle=\frac{|k\rangle+|l\rangle}{\sqrt 2}
	\end{align}
	for $k\neq l$ can be probed via the operator
	\begin{align}\label{eq:SimpleOperator}
		L=|s_{k,l}\rangle\langle s_{k,l}|.
	\end{align}
	This yields the expectation value $\langle s_{k,l}| L|s_{k,l}\rangle=1$ and the bound for classical qudit states $g_{\max}=1/2$.
	The criterion \eqref{eq:Criterion} is satisfied, $\langle s_{k,l}| L|s_{k,l}\rangle>g_{\max}$.
	Therefore, single-particle quantum correlations of the superposition state \eqref{eq:SimpleState} are detected.

	Based on criterion \eqref{eq:Criterion} and motivated by the construction of entanglement monotones in terms of entanglement witnesses \cite{B05,BV06}, we can additionally define the parameter
	\begin{align}\label{eq:GammaParameter}
		\Gamma=\max\left\{
			\frac{\langle L\rangle-g_{\max}}{g_{\max}}
			,0
		\right\}
	\end{align}
	for $g_{\max}\neq0$.
	If quantum correlations are detected via criterion \eqref{eq:Criterion} for an observable $L$, we have $\Gamma>0$.
	In other words, the parameter \eqref{eq:GammaParameter} quantifies the relative strength of the verified quantum correlations.
	For the single-qudit state \eqref{eq:SimpleState} and the test operator \eqref{eq:SimpleOperator}, we get a relative strength of $\Gamma=1$.
	This single-qudit case presented here is certainly a trivial one, but it demonstrates the general proceeding in this section when discussing more complex scenarios.

\subsection{Two particles}\label{subsec:Qudit2}

	Let us consider several examples of bipartite ($N=2$) quantum coherence.
	We especially demonstrate how local and global quantum coherences are affected by entanglement and the exchange symmetry.
	For this reason, we define a number of quantum correlated states $|\psi\rangle$ and then apply the corresponding test operator $L=|\psi\rangle\langle\psi|$, which results in the expectation value $\langle\psi|L|\psi\rangle=1$.
	Then, we list the classical bounds for classical states $g_{\max}^{(s)}$ for distinguishable particles, bosons, and fermions ($s=0,\pm$) using Eqs. \eqref{eq:MultiparticleBoundsTensor}, \eqref{eq:MultiparticleBoundsFermions}, and \eqref{eq:MultiparticleBoundsBosons}, respectively.
	Also, the bounds $g_{\max}^{(\mathrm{sep},s)}$ for separable states, presented in the Appendix, are also given to probe entanglement.
	Finally, we discuss the conclusions in some detail.

	In order to classify different forms of quantum correlated states, we may list characteristic representatives.
	The states to be characterized are
	\begin{align}\label{eq:examples}
	\begin{aligned}
		|\psi_{1}^{(0)}\rangle=& |0\rangle\otimes|s_{1,2}\rangle,
		\\
		|\psi_{1}^{(\pm)}\rangle=& \frac{|0\rangle\otimes|s_{1,2}\rangle\pm|s_{1,2}\rangle\otimes|0\rangle}{\sqrt 2},
		\\
		|\psi_{2}^{(0)}\rangle=& |s_{0,1}\rangle\otimes|s_{2,3}\rangle,
		\\
		|\psi_{2}^{(\pm)}\rangle=& \frac{|s_{0,1}\rangle\otimes|s_{2,3}\rangle\pm|s_{2,3}\rangle\otimes|s_{0,1}\rangle}{\sqrt 2},
		\\
		|\psi_{3}^{(0)}\rangle=& \frac{|0\rangle\otimes|1\rangle+|2\rangle\otimes|3\rangle}{\sqrt 2},
		\\
		|\psi_{3}^{(\pm)}\rangle=& \frac{|0\rangle\otimes|1\rangle\pm|1\rangle\otimes|0\rangle+|2\rangle\otimes|3\rangle\pm|3\rangle\otimes|2\rangle}{2},
	\end{aligned}
	\end{align}
	where the definition \eqref{eq:SimpleState} for the states $|s_{k,l}\rangle$ is used.
	In Table \ref{tab:Bipartite}, we give the bounds for the different concepts of classicality under study.
	In the following, we classify the quantum correlations of the bipartite states \eqref{eq:examples}.
	Based on the parameter $\Gamma$ [Eq. \eqref{eq:GammaParameter}], we specifically focus on the comparison of the impact of entanglement and the exchange symmetry on the quantum coherence.
	Note that for the given examples, a smaller bound implies a larger relative correlation parameter.

\begin{table}[t]
	\caption{
		Bounds for operators $L=|\psi\rangle\langle\psi|$ for the vectors $|\psi\rangle$ in Eq. \eqref{eq:examples}.
		Here, n.a. (not applicable) indicates that the given vector is not in the considered symmetric or antisymmetric subspace.
	}\label{tab:Bipartite}
	\begin{tabular}{cccccc}
		\hline\hline
			\hspace{2ex}Vector\hspace{2ex} &\quad\quad& \hspace{2ex}$g^\mathrm{(0)}_{\max}$\hspace{2ex} & \hspace{2ex}$g^\mathrm{(\pm)}_{\max}$\hspace{2ex} & \hspace{1ex}$g^\mathrm{(sep,0)}_{\max}$\hspace{2ex} & \hspace{2ex}$g^\mathrm{(sep,\pm)}_{\max}$\hspace{2ex}
		\\\hline
			$|\psi_{1}^{(0)}\rangle$ && 1/2 & n.a. & 1 & n.a.
		\\
			$|\psi_{1}^{(\pm)}\rangle$ && 1/4 & 1/2 & 1/2 & 1
		\\
			$|\psi_{2}^{(0)}\rangle$ && 1/4 & n.a. & 1 & n.a.
		\\
			$|\psi_{2}^{(\pm)}\rangle$ && 1/8 & 1/4 & 1/2 & 1
		\\
			$|\psi_{3}^{(0)}\rangle$ && 1/2 & n.a. & 1/2 & n.a.
		\\
			$|\psi_{3}^{(\pm)}\rangle$ && 1/4 & 1/2 & 1/4 & 1/2
		\\
		\hline\hline
	\end{tabular}

\end{table}

	The state $|\psi_{1}^{(0)}\rangle$ is a product state, $\Gamma^{(\mathrm{sep},0)}=0$, and its quantum correlations are therefore completely determined by the nonclassicality ($\Gamma^{(0)}=1>0$) of the single qudit.
	In fact, its form $|\psi_{1}^{(0)}\rangle=|0\rangle\otimes|s_{1,2}\rangle$ indicates that we have only local quantum superposition (see also the single-qudit example in Sec. \ref{subsec:Qudit1}).
	The second state, $|\psi_{1}^{(\pm)}\rangle$, is an (anti)symmetric version of $|\psi_{1}^{(0)}\rangle$, i.e., $|\psi_{1}^{(\pm)}\rangle\cong\Pi_\pm|\psi_{1}^{(0)}\rangle$.
	Because of the exchange symmetry, we can observe an increase of quantum coherence with respect to $\otimes$, $\Gamma^{(0)}=3>1$.
	However, when including the exchange symmetry in the definition of multimode classical states $\mathcal C^{(\pm)}$, we get the same amount of correlations, $\Gamma^{(\pm)}=1$ for $|\psi_{1}^{(\pm)}\rangle$, as we got for the asymmetric state $|\psi_{1}^{(0)}\rangle$.
	In addition, there is no entanglement for $|\psi_{1}^{(\pm)}\rangle$ when considering $\vee$ and $\wedge$, $\Gamma^\mathrm{(\mathrm{sep},\pm)}=0$.
	However, entanglement with respect to $\otimes$ can be observed, $\Gamma^\mathrm{(\mathrm{sep},0)}=1$, for this (anti)symmetric state (see Sec. \ref{subsec:QuestionEntanglement}).

	The state $|\psi_{1}^{(0)}\rangle$ was shown to exhibit only local quantum coherences.
	More specifically, its definition as $|0\rangle\otimes|s_{1,2}\rangle$ suggests that we have one-sided superpositions.
	For comparison, the separable state $|\psi_{2}^{(0)}\rangle=|s_{0,1}\rangle\otimes|s_{2,3}\rangle$ ($\Gamma^\mathrm{(sep,0)}=0$) shows two-sided, local quantum correlations, $\Gamma^{(0)}=3>0$.
	Its (anti)symmetric version $|\psi_{2}^{(\pm)}\rangle$ includes quite strong quantum coherences for the distinguishable and indistinguishable cases, $\Gamma^{(0)}=7$ and $\Gamma^{(\pm)}=3$.
	Again, we either can or cannot observe entanglement of the state $|\psi_{2}^{(\pm)}\rangle$, depending on the choice of the tensor product, $\Gamma^\mathrm{(sep,0)}=1$ and $\Gamma^\mathrm{(sep,\pm)}=0$.

	Finally, we study the states $|\psi_{3}^{(0)}\rangle$ and its projection $|\psi_{3}^{(\pm)}\rangle\cong\Pi_\pm|\psi_{3}^{(0)}\rangle$ onto the subspace of bosons or fermions.
	The quantum coherences of these states are completely determined by entanglement because we have $\Gamma^{(s)}=\Gamma^{(\mathrm{sep},s)}=1>0$ for the applicable values of $s\in\{0,\pm\}$.
	Hence, we can say that this example includes only global quantum superpositions.
	Also, the state $|\psi_{3}^{(\pm)}\rangle$ is entangled regardless of the choice of the tensor product, $\Gamma^{(\mathrm{sep},0)}=3>0$ and $\Gamma^{(\mathrm{sep},\pm)}=1>0$.

	Hence, we have been able to identify different sources and the relative strength of quantum correlations for two qudits (see also Fig. \ref{fig:Concept}).
	We could infer local (one- and two-sided) superpositions as well as global ones.
	The global quantum coherences have been related to the entanglement.
	Moreover, we compared and identified the impact of the exchange symmetry on the quantum coherence.
	For the studied examples, this included the influence on quantum correlations and entanglement from the symmetrization.

\subsection{Mixed, continuous-variable bosonic state}

	To cover the case of a continuous-variable system, $d=\infty$, we may consider the bipartite state
	\begin{align}\label{eq:CVstate}
		|\chi_\kappa\rangle=\sqrt{1-\kappa^2}\sum_{k=0}^\infty \kappa^{k} |k\rangle\otimes|k\rangle,
	\end{align}
	where $0\leq\kappa<1$.
	When the first and second subsystems correspond to two optical modes, this state is termed a two-mode squeezed-vacuum state.
	The state \eqref{eq:CVstate} is symmetric because $|k\rangle\otimes|k\rangle\cong|k\rangle\vee|k\rangle$ holds for all $k$.
	To study a case where the test operator is different from the actual state under study, we may also define the unnormalized vector
	\begin{align}\label{eq:CVvector}
		|\chi\rangle=\sum_{k=0}^\infty |k\rangle\otimes|k\rangle.
	\end{align}
	Thus, for the operator $L=|\chi\rangle\langle\chi|$, we get
	\begin{align}\label{eq:CVbounds}
		g_{\max}^{(0)}=g_{\max}^{(+)}=g_{\max}^{(\mathrm{sep},0)}=1
		\text{ and }
		g_{\max}^{(\mathrm{sep},+)}=2
	\end{align}
	from Sec. \ref{subsect:QuditDetection} and the Appendix.

	Furthermore, let us include a decoherence process as a possible source of loss of coherence \cite{Z03,S05}.
	As an example, we assume that the subsystems are jointly subjected to a fluctuation of a phase $\varphi$, described by the unitary operation $U(\varphi)=\sum_{k=0}^\infty e^{ik\varphi}|k\rangle\langle k|$, which is uniformly distributed over the interval $[-\Delta\varphi/2,+\Delta\varphi/2]$.
	Hence, we have the mixed state
	\begin{align}\label{eq:CVstatemixed}
	\begin{aligned}
		&\varrho_{\Delta\varphi}=
		\int\limits_{-\frac{\Delta\varphi}{2}}^{+\frac{\Delta\varphi}{2}} \frac{d\varphi}{\Delta\varphi}\,
		U(\varphi)^{\otimes 2}|\chi_\kappa\rangle\langle\chi_\kappa|U(\varphi)^{\dag\otimes 2}
		\\
		=&\sum_{k,l=0}^\infty (1-\kappa^2)\kappa^{k+l}\mathrm{sinc}([k-l]\Delta\varphi)
		|k\rangle\langle l|\otimes|k\rangle\langle l|,
	\end{aligned}
	\end{align}
	using $\mathrm{sinc}(x)=\sin(x)/x$ and $X^{\otimes 2}=X\otimes X$.
	For the nondephased case, $\Delta\varphi=0$, we retrieve the pure state in Eq. \eqref{eq:CVstate}, and we get an incoherent state for $\Delta\varphi=\pi$,
	\begin{align}
		\varrho_{\pi}=(1-\kappa^2)\sum_{k=0}^\infty \kappa^{2k} |k\rangle\langle k|\otimes|k\rangle\langle k|.
	\end{align}
	This means $\varrho_{\pi}\in\mathcal I^{(0)}$ and $\varrho_{\pi}\in\mathcal I^{(+)}$.
	Finally, we can compute the expectation value of $L$ \cite{comment3},
	\begin{align}
	\begin{aligned}
		\langle L\rangle=&\langle \chi|\varrho_{\Delta\varphi}|\chi\rangle
		=\int\limits_{-\frac{\Delta\varphi}{2}}^{+\frac{\Delta\varphi}{2}} \frac{d\varphi}{\Delta\varphi}\,\frac{1-\kappa^2}{1+\kappa^2-2\kappa\cos(2\varphi)}
		\\
		=&\frac{2}{\Delta\varphi}\arctan\left(\frac{1+\kappa}{1-\kappa}\tan\left[\frac{\Delta\varphi}{2}\right]\right).
	\end{aligned}
	\end{align}

\begin{figure}[t]
	\includegraphics[width=\columnwidth]{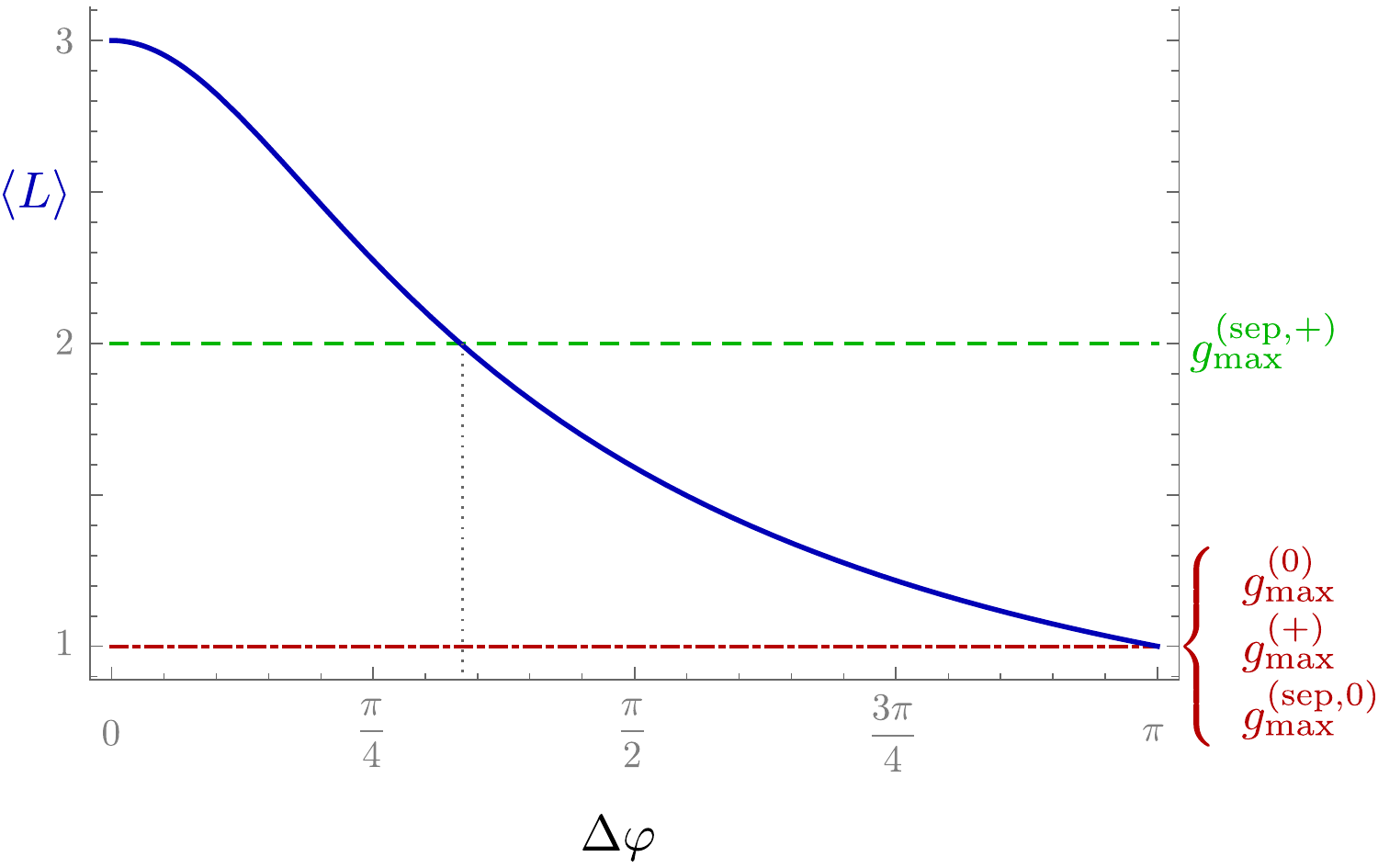}
	\caption{(Color online)
		The expectation value of the test operator $L=|\chi\rangle\langle \chi|$ [Eq. \eqref{eq:CVvector}] for the state $\varrho_{\Delta\varphi}$ [Eq. \eqref{eq:CVstatemixed}] is shown as a function of the dephasing parameter $\Delta\varphi$ and for $\kappa=1/2$.
		The maximal expectation values for the different forms of classical states [Eq. \eqref{eq:CVbounds}] are depicted as horizontal dashed and dot-dashed lines.
		For all $\Delta\varphi<\pi$, we verify quantum coherence for the distinguishable and indistinguishable cases as well as for entanglement based on $\otimes$.
		Bosonic entanglement, using $\vee$, is detected for a dephasing less than the bound which is indicated by the vertical, dotted line.
	}\label{fig:cv}
\end{figure}

	In Fig. \ref{fig:cv}, the dependence of the verified quantum correlations on the dephasing parameter $\Delta\varphi$ is studied for $\kappa=1/2$.
	Quantum coherence is equally well certified for the scenario of distinguishable degrees of freedom and the bosonic case, $\langle L\rangle>g_{\max}^{(0)}=g_{\max}^{(+)}$, for the full interval of dephasing parameters, $0\leq\Delta\varphi<\pi$.
	Also, the entanglement with respect to the standard tensor product follows the same behavior.
	However, bosonic entanglement cannot be detected by this observable from a certain dephasing onwards.
	Since $g_{\max}^{(\mathrm{sep},+)}>g_{\max}^{(\mathrm{sep},0)}$, the detected bosonic entanglement is, in general, weaker than the entanglement based on distinguishable particles.

\subsection{Multiparticle correlations}

	Beyond bipartitions, multiparticle systems allow for the possibility of global quantum correlations on a partial or global scale.
	To study their meanings, let us consider the tripartite examples
	\begin{align}\label{eq:TripartiteStates}
	\begin{aligned}
		|\psi_{4}^{(0)}\rangle=&|0\rangle\otimes\frac{|1\rangle\otimes|2\rangle+|3\rangle\otimes|4\rangle}{\sqrt 2},
		\\
		|\psi_{5}^{(0)}\rangle=&\frac{|0\rangle\otimes|1\rangle\otimes|2\rangle+|3\rangle\otimes|4\rangle\otimes|5\rangle}{\sqrt 2},
		\\
		|\psi_{4}^{(\pm)}\rangle\cong\Pi_\pm&|\psi_{4}^{(\pm)}\rangle,
		\text{ and }
		|\psi_{5}^{(\pm)}\rangle\cong\Pi_\pm|\psi_{5}^{(\pm)}\rangle.
	\end{aligned}
	\end{align}
	Again, the states $|\psi_{n}^{(\pm)}\rangle$ are the (anti)symmetric version of the states $|\psi_{n}^{(0)}\rangle$, $n=4,5$.
	In addition and by construction, the state $|\psi_4^{(0)}\rangle$ is partially separable; the first mode factorized from the joint mode consisting of the second and third particles.
	The state $|\psi_4^{(0)}\rangle$ is a GHZ-type state \cite{GHZ89}, which can be seen by applying local flip operators, $[F_1\otimes F_2\otimes F_3]|\psi_{5}^{(0)}\rangle=(|0\rangle^{\otimes 3}+|1\rangle^{\otimes 3})/\sqrt 2$ \cite{comment1}.

	First, let us focus on the relation between quantum correlations ($|\psi\rangle\notin\mathcal C^{(0)}$), inseparability with respect to full separations ($|\psi\rangle\neq |a\rangle\otimes|b\rangle\otimes|c\rangle$), and full entanglement (i.e., no bipartition is possible).
	Note that quantum correlations are required to have entanglement and that partial inseparability implies full inseparability.
	For the partially separable state $|\psi_{4}^{(0)}\rangle$ using distinguishable degrees of freedom, we find
	\begin{align}\label{eq:PartialGamma}
		\Gamma^{(0)}=1,
		\text{ }
		\Gamma^{(\mathrm{partsep},0)}=0,
		\text{ and }
		\Gamma^{(\mathrm{fullsep},0)}=1,
	\end{align}
	and for the GHZ-type state $|\psi_{5}^{(0)}\rangle$ for distinguishable degrees of freedom, we get
	\begin{align}\label{eq:GHZGamma}
		\Gamma^{(0)}=1,
		\text{ }
		\Gamma^{(\mathrm{partsep},0)}=1,
		\text{ and }
		\Gamma^{(\mathrm{fullsep},0)}=1
	\end{align}
	(see Ref. \cite{SV13} for the used bounds).
	Comparing the correlation parameters \eqref{eq:PartialGamma} for partial separability, we can confirm with our approach that the state exhibits quantum coherence which originates entirely from full inseparability, which was the idea of construction for this state $|\psi_{4}^{(0)}\rangle$ [Eq. \eqref{eq:TripartiteStates}].
	Analogously, we find that for the state $|\psi_{5}^{(0)}\rangle$, its quantum coherence is completely due to partial inseparability, which also results in full inseparability.

	After considering the relation between quantum coherence and different forms of entanglement, let us now focus on the interplay between quantum correlations and the exchange symmetry.
	For the state $|\psi_{4}^{(\pm)}\rangle$, we find
	\begin{align}\label{eq:FermionBosonGamma}
		\Gamma^{(0)}=11
		\text{ and }
		\Gamma^{(\pm)}=1.
	\end{align}
	When we considers that the exchange symmetry is a source of coherent quantum superpositions, it is expected to have stronger quantum correlations than in the case that does not make this assumption.
	This means $\Gamma^{(0)}>\Gamma^{(\pm)}$ [Eq. \eqref{eq:FermionBosonGamma}].
	Note that for the state $|\psi_{5}^{(\pm)}\rangle$, we get the same parameters as those given in Eq. \eqref{eq:FermionBosonGamma}.
	This means that the type of entanglement has no influence on the quantum coherence in this example.

	Like for the proof of principle demonstrated here for the tripartite case, we can study multiparticle scenarios with $N>3$ as well.
	For a large number of particles, $N\gg 1$, one can speak about a macroscopic system \cite{FSDGS17}.
	In particular, inseparability in such a scenario can be present in various, complex forms (see, e.g., Ref. \cite{AFOV08} for an overview and Refs. \cite{HV13,LM13,SSV14,SW17} for recent studies).
	This makes the entanglement certification more sophisticated.
	However, the identification of quantum coherence does not add complexity, and it was formulated in its general form for an $N$-particle system in Sec. \ref{subsect:QuditDetection}.

\section{Conclusions}\label{sec:Conclusions}

	In summary, we studied the detection of quantum coherence to identify the source of quantum interference.
	In particular, we characterized quantum correlations which originate from local quantum superpositions, global quantum superposition, and the exchange symmetry.
	For this reason, different forms of entanglement have been additionally determined in connection with global quantum coherence and in relation to the bosonic and fermionic features of a multiparticle quantum system.
	We applied our approach, formulating measurable quantum correlation criteria, to a compound quantum system which consists of multiple qudits.

	When generalizing the single-particle notion of quantum coherence to a system of multiple indistinguishable particles, we found that there are two possibilities to perform this task.
	Namely, one can include or exclude the (anti)symmetrization requirement in the definition of multimode quantum correlations.
	Naturally, this can lead to different interpretations of the strength of quantum effects.
	However, it was also elaborated that this can be beneficial as it allows us to discern the origin of quantum correlations, which can be dependent on or independent of the exchange symmetry of the joint system.

	Furthermore, we discussed the role of quantum entanglement in connection with quantum coherence.
	The inseparability is one quantum correlation, which is even independent of the single-particle notion of a classical state.
	Again, we were able to infer the impact of the exchange symmetry of indistinguishable particles on this form of global quantum coherence.
	This enabled us to classify effects which originate from different forms of multiparticle entanglement, including instances of partial and full entanglement.

	To access such different types of quantum effects in a measurable form, we formulated witnesses to probe the individual forms of quantum properties.
	Let us point out that this method is based on the same observable for all types of quantum correlations.
	In the case of pure states, we determined the optimal choice of such an witness observable, which is a rank-one operator identical to the state under study.
	Furthermore, we also demonstrated that the general method applies to mixed continuous-variable states too.
	Our collection of comprehensive examples included studies of local (one- and two-sided) and global superposition states in the bipartite and multipartite scenarios of finite-dimensional qudits and an infinite-dimensional case in systems of distinguishable and indistinguishable particles.

	Therefore, we performed a joint characterization of quantum coherence, quantum entanglement, and the exchange symmetry in systems of many indistinguishable particles.
	Applicable criteria have been formulated to probe and discriminate the origins of quantum effects.
	Thus, our general method allows us to classify the composition of quantum correlations, which is useful for the fundamental understanding of the superposition principle as well as for applications in quantum technology.

\begin{acknowledgments}
	The project leading to this application has received funding from the European Union's Horizon 2020 research and innovation program under Grant Agreement No. 665148 (QCUMbER).
\end{acknowledgments}

\appendix*
\section{Entanglement detection}\label{app:EntTests}

	For probing the various forms of entanglement, let us consider the construction of entanglement probes of the form \cite{SV09,SV13,RSV15}
	\begin{align}\label{eq:EntanglementCriterion}
		\langle L\rangle>g_{\max}^{(K\text{-sep})},
	\end{align}
	where $g_{\max}^{(K\text{-sep})}$ is the maximal expectation value of $L$ for $K$-separable states.
	Here, a pure separable state has the general form $\mathbb P[|x_1\rangle\otimes\cdots\otimes|x_K\rangle]$, where $K$ indicates that we consider a $K$ partition and $\mathbb P=\mathbb P^\dag=\mathbb P^2$ is an orthogonal projection operator.
	The problem of finding the bound $g_{\max}^{(K\text{-sep})}$ leads to the so-called separability eigenvalue equations \cite{SV09,SV13,RSV15},
	\begin{align}\label{eq:SeparabilityEigenvalueEquations}
		L_{x_1,\ldots,x_{j-1},x_{j+1},\ldots, x_K}|x_j\rangle
		=
		g\mathbb P_{x_1,\ldots,x_{j-1},x_{j+1},\ldots, x_K}|x_j\rangle,
	\end{align}
	for all $j=1,\ldots,K$ and using the definition ($X=L,\mathbb P$)
	\begin{align}
	\begin{aligned}
		&X_{x_1,\ldots,x_{j-1},x_{j+1},\ldots, x_K}
		\\
		=&\mathrm{tr}_{1}\cdots\mathrm{tr}_{j-1}\mathrm{tr}_{j+1}\cdots\mathrm{tr}_{K}
		\Big[
			X
			\\&\times
			|x_{1}\rangle\langle x_{1}|\otimes\cdots\otimes|x_{j-1}\rangle\langle x_{j-1}|
			\otimes \mathbb I_j
			\\&\otimes
			|x_{j+1}\rangle\langle x_{j+1}|\otimes\cdots\otimes|x_{K}\rangle\langle x_{K}|
		\Big],
	\end{aligned}
	\end{align}
	which defines an operator acting on the $j$th party and which depends on the vectors $|x_i\rangle$ for $i\neq j$;
	$\mathbb I_j$ is the identity operator of the $j$th subsystem.
	The separability eigenvalues $g$ in the coupled set of equations \eqref{eq:SeparabilityEigenvalueEquations} yield the bound $g_{\max}^{\rm (sep)}$ for the entanglement condition \eqref{eq:EntanglementCriterion} as
	\begin{align}
		g_{\max}^{(K\text{-sep})}=\max\{ g: \text{$g$ solves Eqs. \eqref{eq:SeparabilityEigenvalueEquations}}\}
	\end{align}
	(cf. \cite{SV09,SV13,RSV15}).
	Let us also mention that we assume that $[L,\mathbb P]=0$ to ensure that $L$ acts on the subspace projected by $\mathbb P$, and we assume that $\mathbb P$ can be either the identity $\mathbb I$ for distinguishable particles or the permutation operators in Eq.~\eqref{eq:PermutationOperator} for indistinguishable ones.

	For self-consistency, we list the known solutions which are relevant for this work.
	Especially, we consider operators of the form
	\begin{align}\label{eq:EntanglementTestOperator}
		L=|\psi\rangle\langle\psi|.
	\end{align}
	The required commutator relation $[L,\mathbb P]=0$ and the idempotence $\mathbb P^2=\mathbb P$ imply that $\mathbb P|\psi\rangle=|\psi\rangle$.
	For instance, we can write for the bipartite scenario of distinguishable degrees of freedom (i.e., $\mathbb P=\mathbb I$) that
	\begin{align}
		|\psi\rangle=\sum_{k=0}^{d-1} \lambda_{k}^{(0)} |e_k\rangle\otimes|f_k\rangle
	\end{align}
	which is the Schmidt decomposition of $|\psi\rangle\in\mathcal H^{\otimes 2}$ \cite{NC00},
	that is the singular-value decomposition of the coefficient matrix $M=(\psi_{k,l})_{k,l=0}^{d-1}$ of the vector $|\psi\rangle=\sum_{k,l}\psi_{k,l} |k\rangle\otimes|l\rangle$ \cite{HJ13}.
	It was shown in Ref. \cite{SV09} that the maximal separability eigenvalue is given by
	\begin{align}
		g_{\max}^{\rm (sep,0)}=\max_k(\lambda_k^{(0)})^2.
	\end{align}

	In the scenario of indistinguishable particles, the coefficient matrix $M$ is symmetric for bosons ($M=M^\mathrm{T}$) and antisymmetric for fermions ($M=-M^\mathrm{T}$).
	This means $\mathbb P=\Pi_+$ for bosons and $\mathbb P=\Pi_-$ for fermions.
	Using the corresponding matrix decompositions for those cases \cite{HJ13}, the vector $|\psi\rangle$ can be written for the bosonic case as
	\begin{align}
		|\psi\rangle=\sum_{k=0}^{d-1} \lambda_k^{(+)} |e_k\rangle\otimes|e_k\rangle.
	\end{align}
	For fermions, the so-called Slater decomposition is similarly obtained,
	\begin{align}
		|\psi\rangle=\sum_{k=0}^{\lfloor d/2\rfloor} \lambda_k^{(-)} \left(
			|e_{2k}\rangle\otimes|e_{2k+1}\rangle-|e_{2k+1}\rangle\otimes|e_{2k}\rangle
		\right).
	\end{align}
	Finally, we get the maximal separability eigenvalues,
	\begin{align}
	\begin{aligned}
		g_{\max}^{\rm (sep,+)}=&\max_{k>l} \big((\lambda_k^{(+)})^2+(\lambda_l^{(+)})^2\big),\\
		g_{\max}^{\rm (sep,-)}=&\max_{k} \big(2(\lambda_k^{(-)})^2\big),
	\end{aligned}
	\end{align}
	from Ref. \cite{RSV15}.



\begin{thebibliography}{99}
\bibitem{comment2}
		Note that we ignore global phases; that is, we identify $|c\rangle$ with $|c_\varphi\rangle=e^{i\varphi}|c\rangle$ for all $0\leq \varphi<2\pi$ since they do not play a role in the density operator, $|c\rangle\langle c|=|c_\varphi\rangle\langle c_\varphi|$.
\bibitem{comment3}
		We apply the integral formula $\int dx\,(b+c\cos[ax])^{-1}=(2/a)(b^2-c^2)^{-1/2}\arctan[(b-c)(b^2-c^2)^{-1/2}\tan[ax/2]]$, where $b^2>c^2$.
\bibitem{comment1}
		The flip operators act as follows:
		$F_1$ exchanges $|3\rangle$ with $|1\rangle$,
		$F_2$ exchanges $|2\rangle$ with $|0\rangle$ and $|4\rangle$ with $|1\rangle$,
		and $F_3$ exchanges $|2\rangle$ with $|0\rangle$ and $|5\rangle$ with $|1\rangle$.
		All other basis elements are mapped onto themselves.
\end{thebibliography}
\end{document}